\documentclass[amsmath,amssymb, reprint]{revtex4-1}
\usepackage{braket}
\usepackage{graphicx,epstopdf,bm}
\usepackage{natbib}

\begin{document}

\title{Direct measurement of the density matrix of a quantum system}


\author{G.S. Thekkadath}
\author{L. Giner}
\author{Y. Chalich}
\author{M.J. Horton}
\author{J. Banker}
\author{J.S. Lundeen}
\affiliation{Department of Physics and Max Planck Centre for Extreme and Quantum Photonics,
University of Ottawa, 25 Templeton Street, Ottawa, Ontario K1N 6N5, Canada}



\begin{abstract}
One drawback of conventional quantum state tomography is that it does not readily provide access to single density matrix elements, since it requires a global reconstruction. Here we experimentally demonstrate a scheme that can be used to directly measure individual density matrix elements of general quantum states. The scheme relies on measuring a sequence of three observables, each complementary to the last. The first two measurements are made weak to minimize the disturbance they cause to the state, while the final measurement is strong. We perform this joint measurement on polarized photons in pure and mixed states to directly measure their density matrix. The weak measurements are achieved using two walk-off crystals, each inducing a polarization-dependent spatial shift that couples the spatial and polarization degrees of freedom of the photons. This direct measurement method provides an operational meaning to the density matrix and promises to be especially useful for large dimensional states.
\end{abstract}

\maketitle

Shortly after the inception of the quantum state, Pauli questioned its measurability, and in particular, whether or not a wave function can
be obtained from position and momentum measurements~\cite{Pauli1980}. 
This question, now referred to as the Pauli problem, draws on concepts such as complementarity and measurement in an attempt to demystify the physical significance of the quantum state. Indeed, the task of determining a quantum state is a central issue in quantum physics due to both its foundational and practical implications. For instance, a method to verify the production of complicated states is desirable in quantum information and quantum metrology applications. Moreover, since a state fully characterizes a system, any possible measurement outcome can be predicted once the state is determined. 

A wave function describes a quantum system that can
be isolated from its environment, meaning the two are non-interacting and the system is in a pure state.
More generally, open quantum
systems can interact with their environment and the two can become entangled. In such
cases, or even in the presence of classical noise, the system is in 
a statistical mixture of states (\textit{i.e.} mixed state), and one requires 
a density matrix to fully describe the quantum system.
In fact, some regard the density matrix as more fundamental 
than the wave function because of its generality and its relationship to classical measurement
theory~\cite{hardy2001quantum}.

The standard way of measuring the density matrix is by using quantum state tomography (QST). In QST, one performs
an often overcomplete set of measurements in incompatible bases on identically prepared copies of the state. Then, one fits a candidate state to the measurement results with the help of a reconstruction algorithm~\footnote{Some reconstruction algorithms restrictively fit the measurement results to reconstruct only physical (\textit{i.e.} positive semi-definite and normalized) states.}. Many efforts have been made to optimize 
QST~\cite{wooters1989optimal,renes2004symmetric,durt2010mutually,bent2015experimental}, but the scalability of the experimental apparatus and the complexity of the
reconstruction algorithm renders the task increasingly difficult for large dimensional
systems. In addition, since QST requires a global reconstruction, it does not 
provide direct access to coherences (\textit{i.e.} off-diagonal elements), which are of particular interest in quantum physics.


Some recent work has focused on developing
a direct approach to measuring quantum states~\cite{lundeen2011direct,lundeen2012procedure,bamber2014observing,salvail2013full,mirhosseini2014compressive,malik2014direct,fischbach2012quantum,di2013sequential,wu2013state,bolduc2016direct,shi2015scan,vallone2016strong}. Defining features of direct methods are that they can determine the state without complicated computations, and they can do so locally, \textit{i.e.} at the location of the measurement probe. For example,
direct measurement of the wave function has been achieved by
performing a sequence consisting of a weak and strong measurement
of complementary variables (\textit{e.g.} position and momentum)~\cite{lundeen2011direct}. In the sub-ensemble of trials for which the strong measurement results in a particular outcome (i.e. ``post-selection"), the average weak measurement outcome is a complex number
known as the weak value~\cite{aharonov1988how,aharonov1990properties}. The weak value is a concept that has proven to be useful in addressing fundamental
questions in quantum physics~\cite{resch2004experimental,lundeen2009experimental,yokota2009direct,palacios2010experimental,goggin2011violation,dressel2011experimental,steinberg1995much,kocsis2011observing,wiseman2002weak,piacentini2016experiment},
even beyond optics~\cite{demonstration2013shomroni}. By foregoing post-selection, previous work~\cite{bamber2014observing,salvail2013full} generalized the direct wave function measurement scheme to 
measure mixed quantum states. However, their method still does
not provide direct access to individual density matrix elements. Ref.~\cite{lundeen2012procedure} proposes a way to do this by performing an additional complementary measurement after the wave function measurement sequence: The second measurement serves as a phase reference and enables the first and last measurements to probe the coherence between any two chosen states in some basis. On top of its applications, a direct measurement method provides
an operational meaning to the density matrix in terms of a sequence of three complementary measurements. 

In this Letter, we experimentally demonstrate the method proposed in Ref.~\cite{lundeen2012procedure} by directly
measuring any chosen element of a density matrix $\bm{\rho}_\mathcal{S}$ of a system $\mathcal{S}$.
By repeating this for each element, we then measure the entire density
matrix, thereby completely determining the state of the system. At the center
of the method is a sequence of incompatible measurements~\cite{piacentini2015measuring,suzuki2016observation}.
In order for these measurements not to disrupt each other, they are made
weak, a concept that we outline now (for a review, see ~\cite{colloquium2014dressel}). Suppose one wishes to measure the observable $\bm{C}$. In
von Neumann's model of measurement, the measured system $\mathcal{S}$ is coupled
to a separate ``pointer'' system $\mathcal{P}$ whose wave function is initially
centered at some position and has a width $\sigma$. This
coupling proportionally shifts the position of the pointer by the value
of $\bm{C}$ as described by the unitary translation $\bm{U}=\exp({-i\delta\bm{C}\bm{p}/\hbar})$,
where $\bm{p}$ is the pointer momentum operator and $\delta$ is
strength of the interaction. After the coupling, the pointer position $q$
is measured. On a trial by trial basis, if $\delta\gg\sigma$,
the pointer position will be shifted by $\Delta q \approx \delta c$ and thus will
indicate that the result of the measurement of $\bm{C}$ is $c$.

In contrast, in weak
measurement $\delta \ll\sigma$, and the measurement result is ambiguous
since it falls within the original position distribution of the pointer.
However, this does have a benefit: The small interaction leaves the
measured system relatively undisturbed and thus it can subsequently
be measured again~\cite{ispen2015disturbance}. By repeating the
weak measurement on an ensemble of systems and averaging, the shift
of the pointer can be found unambiguously. This average shift is called
the ``weak average'' $\braket{\bm{C}}_\mathcal{S}$ and is equal to the expectation value of a
conventional (\textit{i.e. }``strong'') measurement: $\braket{\bm{C}}{_\mathcal{S}}= \mathrm{Tr}_\mathcal{S}[\bm{C}\bm{\rho}_\mathcal{S}]$~\cite{lundeen2012procedure}.
This differs from the weak value normally encountered in that
there is no post-selection.

Unlike in strong measurement, $\bm{C}$
can be non-Hermitian. This is the case when $\bm{C}$ is the product of incompatible
observables which normally disturb each other. Consequently,
it is possible for the weak average to be complex. What does
this imply? Both the position $\bm{q}$ and momentum $\bm{p}$ of the pointer will be
shifted according to
$\braket{\bm{C}}_\mathcal{S} = \frac{1}{\delta}\bm{\braket{\mathrm{a}}}_\mathcal{P}$,
where $\bm{\mathrm{a}}=\bm{q}+i2\sigma^2\bm{p}/\hbar$ is the standard harmonic oscillator lowering operator scaled by $2\sigma$~\cite{lundeen2005practical}.
The real part and imaginary parts of the weak average are
proportional to the average shift of the pointer's position and momentum,
respectively. 

Consider the weak measurement of an observable composed of the following
three incompatible projectors: 
\begin{equation}
\bm{\Pi}_{a_{i}a_{j}}=\bm{\pi}_{a_{j}}\bm{\pi}_{b_{0}}\bm{\pi}_{a_{i}},
\label{eqn:weakprojector}
\end{equation}
where $\bm{\pi}_{a_{i}}=\ket{a_{i}}\bra{a_{i}}$ and $\bm{\pi}_{b_{0}}=\ket{b_{0}}\bra{b_{0}}$, which are composed of eigenstates of the observables $\bm{A}$ and $\bm{B}$, respectively. These are maximally incompatible, or ``complementary'', in the sense that $\left|\braket{a_{i}|b_{0}}\right|=1/\sqrt{d}$
for a $d$-dimensional Hilbert space. In the basis of the eigenstates of $\bm{A}$, a density matrix element is given by $\rho_\mathcal{S}(i,j)=\braket{a_{i}|\bm{\rho}_\mathcal{S}|a_{j}}$. This can be connected to the weak average of the measurement sequence in Eq.~\ref{eqn:weakprojector}:
\begin{equation}
\braket{\bm{\Pi}_{a_{i}a_{j}}}_{\mathcal{S}}= \mathrm{Tr}_\mathcal{S} \left[ \bm{\pi}_{a_{j}}\bm{\pi}_{b_{0}}\bm{\pi}_{a_{i}}\bm{\rho}_\mathcal{S} \right]=\rho_\mathcal{S}(i,j)/d.\label{eqn:weak_density}
\end{equation}
In fact, one can replace the weak measurement of the last projector $\bm{\pi}_{a_{j}}$ by a strong measurement without affecting the weak average~\cite{lundeen2012procedure}, thereby reducing the complexity of the measurement apparatus. Thus any density matrix element can be obtained by selecting the first and last projectors in the measurement sequence. Whichever state $\ket{b_0}$ that is	 chosen for the middle complementary projector serves as a reference for zero phase in the density matrix by fixing $\theta = 0$ for all $a$ in $\braket{a|b_0} = \exp{(i\theta)}/\sqrt{d}$. As such, it should remain fixed.

\begin{figure}
\centering \includegraphics[width=0.5\textwidth]{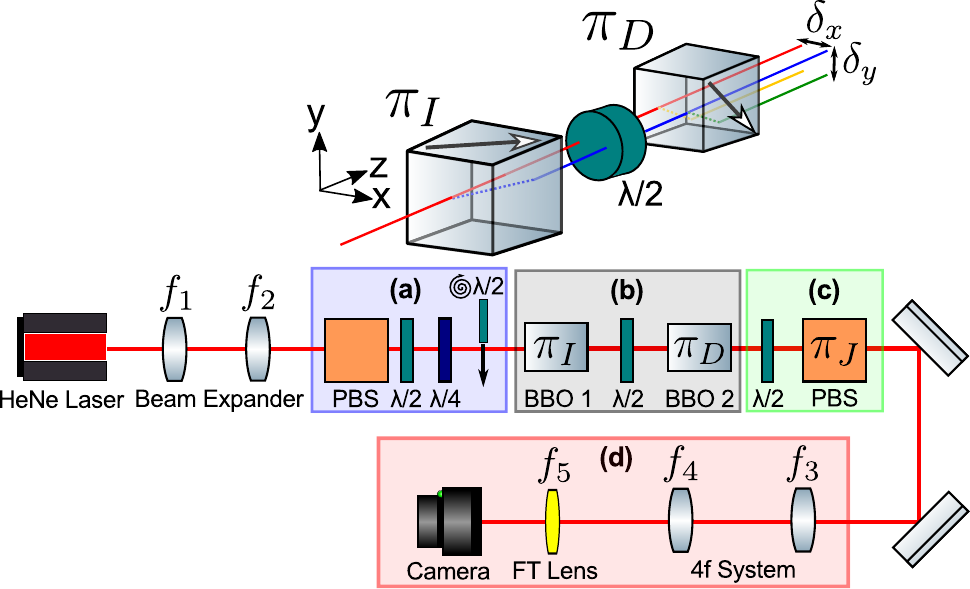} 
\caption{Direct measurement experimental setup. \textbf{(a)} \textit{State preparation}: We use a HeNe laser as a
source of photons. The photon polarization state is set using a half-wave
plate ($\lambda/2$) and a quarter-wave plate ($\lambda/4$). A spinning $\lambda / 2$ is included when generating mixed states. \textbf{(b)} \textit{Weak measurements}: Two subsequent weak measurements $\bm{\pi}_I$ and $\bm{\pi}_D$ 
are performed, each with a walk-off crystal (BBO) that couples the polarization to a spatial degree of freedom, x or y, our measurement pointers. Note that $\delta_x = \delta_y \equiv \delta$.  \textbf{(c)} \textit{Strong measurement}:
The final measurement $\bm{\pi}_J$ is performed by a polarizing beam splitter (PBS),
and the projection direction J is set by a $\lambda/2$.  \textbf{(d)} \textit{Imaging}:
A 4f arrangement of lenses forms an image of the crystal plane onto a
camera allowing us to measure pointer positions. An additional Fourier transform (FT) lens, either spherical or cylindrical, is used to measure pointer momenta.}
\label{fig:setup} 
\end{figure}

The experimental setup is shown in Fig.~\ref{fig:setup}. We demonstrate
the technique by directly measuring the density matrix of a photon
polarization state. This is possibly the simplest system for a demonstration,
but it is also an important one since it can act as a qubit from which
larger and more complicated quantum states can be constructed, such as in quantum
computing. A HeNe laser at 633 nm is sent through a polarizing beam
splitter (PBS) to ensure it is polarized. We treat the bright polarized
beam as a source of a large number of identically prepared polarized photons.
Instead of using a separate system, 
we  use the
$x$ and $y$ transverse spatial distributions of the photons as pointers. Both
are Gaussian with widths $\sigma = 250~\mu$m (830~$\mu$m FWHM)
that are set using a telescopic arrangement of two convex lenses ($f_1$ = 50~mm and $f_2$ = 100~mm). We set the photon polarization
state $\bm{\rho}_\mathcal{S}$ using a half-wave plate ($\lambda/2$) and a quarter-wave
plate ($\lambda/4$). 

A weak measurement of polarization is implemented by coupling the
polarization degree of freedom (our system) to a spatial one (a pointer).
This is accomplished with a walk-off crystal (beta barium borate,
BBO) that shifts the $\ket{I}$ polarization component along $x$ by $\delta=176$~$\mu$m. If $\delta\gg\sigma$ this implements
a strong measurement of $\bm{\pi}_{I}=\ket{I}\bra{I}$ since the
photon position unambiguously determines $\bm{\pi}_{I}$. If $\delta\ll\sigma$
this is a weak measurement of $\bm{\pi}_{I}$~\footnote{A walk-off crystal is used for weak measurement of polarization
in, for example, Refs.~\cite{salvail2013full,ritchie1991realization}}. In our demonstration, we find each of the four polarization density
matrix elements $\rho_\mathcal{S}(I,J)$ by measuring the three projector observable
$\bm{\pi}_{J}\bm{\pi}_{D}\bm{\pi}_{I}$ where either $I$ or $J$
can be horizontal $H$ or vertical $V$ polarization and $\ket{D}=(\ket{H}+\ket{V})/\sqrt{2}$
is a complementary state, the diagonal polarization.

Coupling a joint observable $\bm{E}\bm{F}$ such as $\bm{\pi}_{D}\bm{\pi}_{I}$ to a single pointer is challenging for photons.
Instead, we follow a strategy commonly used for joint strong measurements
(\textit{e.g.} those in Bell's inequalities) in which one independently measures
single observables and then evaluates correlations between the independent
results. In von Neumann's model, this corresponds to having two independent
pointers so that $\braket{\bm{E}\bm{F}}_\mathcal{S} =\left(\frac{1}{\delta}\right)^{2}\braket{\bm{q}_{E}\bm{q}_{F}}_\mathcal{P}$,
where $\bm{q}_{m}$ is the position of the $m=E,F$ pointer. In the weak measurement analog, proposed in Ref.~\cite{lundeen2005practical}, one replaces $\bm{q}_{m}$ by $\bm{\mathrm{a}}_{m}$, and so $\braket{\bm{E}\bm{F}}_\mathcal{S} = \left(\frac{1}{\delta}\right)^{2}\braket{\bm{\mathrm{a}}_{E}\bm{\mathrm{a}}_{F}}_\mathcal{P}$~\cite{mitchison2008weak,lundeen2012procedure}.
Thus we can couple $\bm{\pi}_{I}$ and $\bm{\pi}_{D}$ to
separate pointers and then measure correlations between the momenta and positions of these pointers to find the weak average. The final measurement in the sequence $\bm{\pi}_J$ is strong and so the full joint expectation value is
\begin{equation}
\braket{\bm{\Pi}_{IJ}}_{\mathcal{S}} = \left(\frac{1}{\delta}\right)^2\mathrm{Tr}_\mathcal{T}[\bm{\pi}_J \bm{\mathrm{a}}_D\bm{\mathrm{a}}_I \bm{\rho}_\mathcal{T}] = \rho_\mathcal{S}(I,J)/2,
\label{eqn:joint_pointers}
\end{equation}
where $\mathcal{T} = \mathcal{S} \otimes \mathcal{P}$ indicates the total Hilbert space, combining the pointers and the system ($d=2$).

In our experiment, we
conduct two independent weak measurements by sequentially introducing
two walk-off crystals in the beam path (see Ref.~\cite{sup} for alignment
procedure). The first measures $\bm{\pi}_{I}$ by inducing a displacement
$\delta$ along $x$. Combined with a $\lambda/2$ at $22.5^\circ$, the second
crystal induces a displacement $\delta$ along $y$, measuring $\bm{\pi}_D$.
The last projector $\bm{\pi}_J$,
the strong measurement, is implemented by a second $\lambda/2$
and a PBS where the $\lambda/2$ is used to choose the projected
state $J=H,V$, \textit{i.e.} a J polarizer.

The lowering operators in the total pointer-system expectation value in Eq.~\ref{eqn:joint_pointers} imply the measurement of positions and momenta of the photons. Experimentally, we measure quantities such as the probability that a photon is transmitted through the final J polarizer, and also has horizontal position $x$ and vertical position $y$, \textit{i.e.} $\mathrm{Prob}(x,y,J)$~\footnote{We note that this is not the same as a post-selection since we do not re-normalize the state after passing through the J polarizer, \textit{i.e.} the probability is not conditional on $\bm{\pi}_J$. $\mathrm{Prob}(x,y,J)$ is one value of the total normalized probability distribution.}. From this, we can find expectation values such as $\iint xy\mathrm{Prob}(x,y,J)dxdy \equiv \braket{\bm{x}\bm{y}}_{\mathcal{P},J}$ (see Ref.~\cite{sup} for an example). Then the density matrix elements can be directly related to  
the joint position $(\bm{x},\bm{y})$ and momentum $(\bm{p}_x,\bm{p}_y)$ expectation values of the pointer state: 
\begin{equation}
\begin{aligned}\
\mathrm{Re}[\rho_\mathcal{S}(I,J)] & =\frac{2}{\delta^2}\left(\braket{\bm{x}_I\bm{y}_D}_{\mathcal{P},J}-\frac{\sigma^2}{\sigma_p^2}\braket{\bm{p}_{xI}\bm{p}_{yD}}_{\mathcal{P},J}\right), \\
\mathrm{Im}[\rho_\mathcal{S}(I,J)] & =\frac{2}{\delta^2}\frac{\sigma}{\sigma_p}\left(\braket{\bm{p}_{xI}\bm{y}_D}_{\mathcal{P},J}+\braket{\bm{x}_I\bm{p}_{yD}}_{\mathcal{P},J}\right).
\end{aligned}
\label{eqn:pointer_expec}
\end{equation}
Eq.~\ref{eqn:pointer_expec} is expressed using $\sigma$ and $\sigma_p$ where $\sigma\sigma_{p} = \hbar/2$ to explicitly remove the unit dependence of position and momentum. The subscript $I$ in \textit{e.g.} $\braket{\bm{x}_I\bm{y}_D}$ indicates the projector $\bm{\pi}_I$ is coupled to the $\bm{x}$ pointer.

We measure the four joint expectation values in Eq.~\ref{eqn:pointer_expec} one at a time using a camera  (CMOS sensor with resolution 2560x1920 and pixel side length of 2.2~$\mu$m). The position
expectation value $\braket{\bm{x}\bm{y}}$ of the pointer state is obtained
using two convex lenses ($f_3$~=~1000~mm and $f_4$~=~1200~mm) in a 4f arrangement that images the crystal plane onto the camera. The momentum expectation value $\braket{\bm{p}_{x}\bm{p}_{y}}$ is obtained by adding a spherical lens ($f_5$~=~1000~mm) one focal length from the camera. We replace the spherical lens with a cylindrical one (also $f_5$) to take a one-dimensional Fourier transform of the pointer states and measure the expectation values $\braket{\bm{p}_{x}\bm{y}}$
and $\braket{\bm{p}_{y}\bm{x}}$ by rotating the axis cylindrical
lens. In order to obtain each and every density matrix element, we repeat these four measurements for all combinations of $(I,J)$.

\begin{figure}
\centering \includegraphics[width=0.5\textwidth]{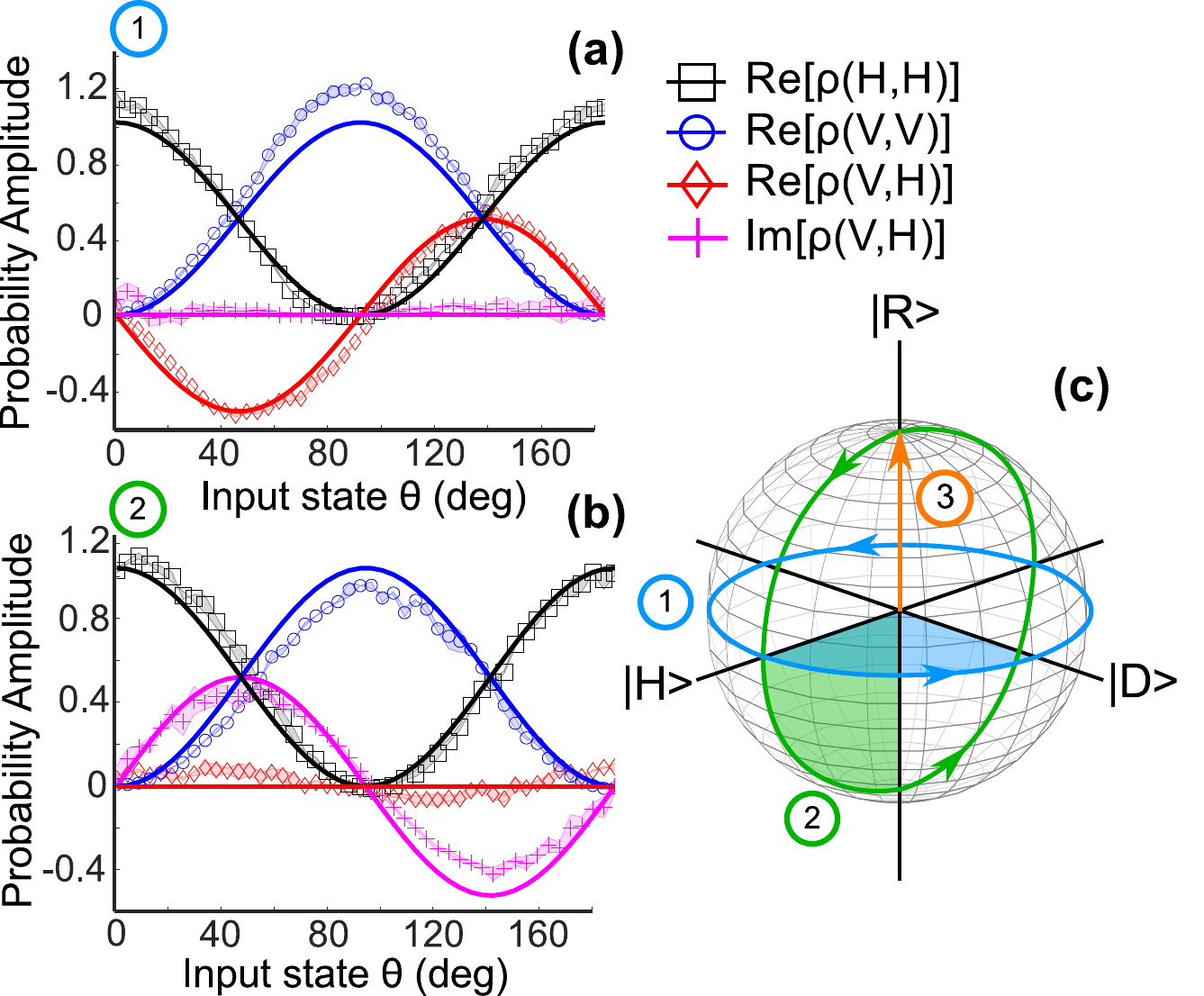} \caption{Direct measurement of density matrix elements for pure polarization states. \textbf{(a)} and \textbf{(b)} are the density matrix elements along path 1 and 2 in the Poincar\'e
sphere, respectively. The bold lines are the theoretical
matrix elements given by Eq.~\ref{eqn:pure_theory}, while the markers
are data points. The shaded region in these plots represents one standard deviation from averaging over ten trials, and is mostly smaller than the size of the markers. \textbf{(c)} Poincar\'e sphere. Path 3 corresponds to the measurement of mixed states, shown in Fig.~\ref{fig:mixed_states}. The shaded regions indicate an interval of $\Delta\theta = 45^\circ$ to help the reader link the paths to the $\theta$ axes in (a) and (b).}
\label{fig:data_pure} 
\end{figure}

\begin{figure*}
\centering 
\includegraphics[width=1.0\textwidth]{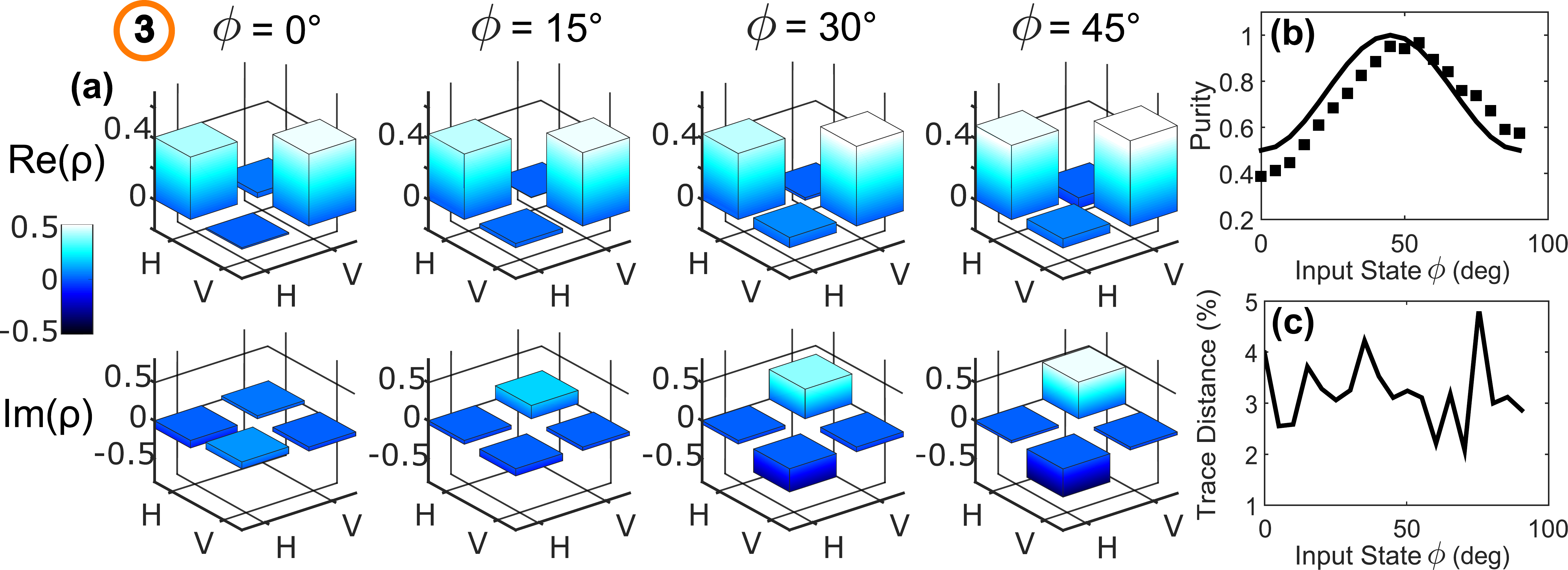} 
\caption{Direct measurement of mixed states. \textbf{(a)} Measured density matrices. The color is proportional to the measured probability amplitudes.  \textbf{(b)} States with various degrees of purity $\mathrm{Tr}[\bm{\rho}^2]$ can be generated by varying the fast axis angle
$\phi$ of a $\lambda/4$, as shown in Eq.~\ref{eqn:mixed_state}. The bold line is the theory while the markers are data points.
The states follow path 3 in the Poincar\'e sphere shown in Fig.~\ref{fig:data_pure}. We do not show statistical uncertainties as they are smaller than the markers. \textbf{(c)} The trace distance is half the Euclidean distance between the measured and theory states on the Poincar\'e sphere, and is always less than 0.049 (\textit{i.e.}~4.9~\%).}
\label{fig:mixed_states} 
\end{figure*}

First, we measure the density matrix elements of the pure state
$\ket{\psi}=\cos{\theta}\ket{H}-\sin{\theta}e^{i\alpha\pi/2}\ket{V}$:
\begin{equation}
\bm{\rho}=\ket{\psi}\bra{\psi}=\begin{pmatrix}\cos^{2}{\theta} & -e^{-i\alpha\pi/2}\cos{\theta}\sin{\theta}\\
-e^{i\alpha\pi/2}\cos{\theta}\sin{\theta} & \sin^{2}{\theta}
\end{pmatrix}.
\label{eqn:pure_theory}
\end{equation}
Fig.~\ref{fig:data_pure}a shows density matrix elements along path
1 in the Poincare sphere, which is traced by setting $\alpha=0$ (\textit{i.e.}
removing the $\lambda/4$) and varying the fast-axis of the $\lambda/2$
such that $\theta\in[0,180^{\circ}]$. Fig.~\ref{fig:data_pure}b shows
the same density matrix elements along path 2, which is by traced
by setting $\alpha=-1$ (\textit{i.e.} $\lambda / 4$ fast axis at $-\pi/2$) and again varying the fast-axis of the $\lambda/2$
such that $\theta\in[0,180^{\circ}]$.
As can be seen, the measured density matrix elements closely follow the theory curve. Deviations from the curve (\textit{e.g.} near $\theta = 90^\circ$) are likely due to imperfections in the wave plates, which can introduce systematic errors both when preparing the polarization state and aligning the BBO crystals.

Next, we generate mixed states by creating a incoherent combination of pure states. This is achieved by introducing a spinning $\lambda/2$ in the preparation stage. This $\lambda/2$ rotates sufficiently fast such that over the exposure
time of the camera, the measured result contains contributions from many polarization states~\cite{sup}. In particular, we produce
\begin{equation} \bm{\rho} =
\begin{pmatrix}1/2 & i\sin{\phi}\cos{\phi}\\
 -i\sin{\phi}\cos{\phi} & 1/2
 \end{pmatrix},
\label{eqn:mixed_state}
\end{equation}
where $\phi$ is the angle between horizontal and
the fast axis of the $\lambda/4$. We generate a series of such mixed states (see Fig.~\ref{fig:mixed_states}b) and vary their purity $\mathrm{Tr}[\bm{\rho}^{2}]$ between 1/2 and 1 by adjusting $\phi$. This corresponds to path 3 in the Poincar\'e sphere, as shown in Fig.~\ref{fig:data_pure}c. To measure the accuracy
of our measured density matrices, we compute the trace distance $\left| \mathrm{Tr}\left[\sqrt{(\bm{\beta}-\bm{\rho})^{\dagger}(\bm{\beta}-\bm{\rho})}\right] \right|/2$
($\bm{\beta}$ is the measured state) which is shown in Fig.~\ref{fig:mixed_states}c.
The trace distance can be interpreted as a measure of the maximum probability of distinguishing between two states,  $\bm{\rho}$ and $\bm{\beta}$, with an optimal measurement. For our results, this probability is always less than 4.9~\%. We also note that the measured density matrix may not be positive semi-definite due to measurement uncertainties. Consequently, if one requires a positive semi-definite matrix, one would need to employ additional algorithms such as a maximum-likelihood estimation.

To summarize, we directly measure the density matrix elements of photons in both
pure and mixed polarization states using three
sequential measurements, each complementary to the last. The first two measurements
are weak to minimize their disturbance on the state, while the last measurement is
strong. The average joint result of this measurement sequence gives any chosen density matrix element, and hence,
can be used to operationally define the density matrix. 

We anticipate that this method will be of use in practical applications.  Since the last measurement can be weak, it could function as a non-invasive probe to determine a quantum state \textit{in situ}, such as during a quantum computation or molecular evolution. Moreover, one could envisage directly observing global properties of a state, such as the existence of non-classical correlations~\cite{suzuki2016observation}, by measuring coherences or entanglement witnesses with our method. Lastly, 
direct measurement has already proven to be efficient for measuring large
dimensional pure states in
various physical systems~\cite{malik2014direct,bolduc2016direct,shi2015scan}.
Quantum state tomography typically requires $\mathcal{O}(d^2)$ measurements in $\mathcal{O}(d)$  bases and finds the full density matrix at once. Thus as \(d\) increases, the experimental procedure and reconstruction algorithm become increasingly complicated. In contrast, our direct measurement method requires three measurements in only two bases to determine any chosen density matrix element regardless of the system dimension \(d\). Consequently, in systems with large \(d\) the method is an attractive alternative to tomography as a way to locally characterize a potentially mixed quantum state.

\begin{acknowledgments}
This work was supported by the Canada Research Chairs (CRC) Program, the Natural Sciences and
Engineering Research Council (NSERC), and
Excellence Research Chairs (CERC) Program.
\end{acknowledgments}

\bibliographystyle{apsrev4-1}
\bibliography{refs}

\end{document}


\section{Supplementary Material}

\subsection{Example of pointer state expectation value}

Here we give an example of a pointer state expectation value for the specific case of $(I,J)=(H,H)$. In this case, the measurement sequence is $\bm{\pi}_H\bm{\pi}_D\bm{\pi}_H$ where \textit{e.g.} $\bm{\pi}_H = \ket{H}\bra{H}$. We assume that the polarization state is pure for simplicity.

The total state $\ket{\Psi} \in \mathcal{H}_\mathcal{T}$ describing the photons before the measurement
is the product state of their spatial $\ket{\chi} \in \mathcal{H}_\mathcal{P}$
and polarization $\ket{\psi} \in \mathcal{H}_{\mathcal{S}}$ degrees of freedom: 
\begin{equation}
\ket{\Psi}=\ket{\chi}\otimes\ket{\psi}=\ket{\chi}\otimes \left( a\ket{H}+b\ket{V} \right).
\end{equation}
The spatial degree of freedom is used as a pointer state $\ket{\chi}=\ket{\chi_{x}}\otimes\ket{\chi_{y}}$ and
is determined by the Gaussian intensity profile of the HeNe laser,
namely 
\begin{equation}
\braket{\zeta|\chi_{\zeta}} \equiv \chi_{\zeta}(\zeta)=(\frac{1}{\sqrt{2\pi}\sigma_{\zeta}})^{1/2}e^{-\frac{\zeta^{2}}{4\sigma_{\zeta}^{2}}}\label{eqn:pointer_pos}
\end{equation}
for $\zeta\rightarrow\left\{ x,y\right\} $. We assume the pointers
are centered at zero before any interaction with the birefringent
crystals. 

The first walk-off crystal implements the weak projector $\bm{\pi}_H$, and hence the
$x$ pointer state $\ket{\chi_{x}}$ associated with the $\ket{H}$ polarization
state is shifted by $\delta_{x}$, which we describe as some unitary
transformation $\bm{U}_{1}$: 
\begin{equation}
\braket{\vec{r}|\bm{U}_{1}|\Psi}=a\chi_{x}(x-\delta_{x})\chi_{y}(y)\ket{H}+b\chi_{x}(x)\chi_{y}(y)\ket{V}
\label{eqn:after_bbo1}
\end{equation}
where $\vec{r}=(x,y)$. We can express the result above in the diagonal basis using $\ket{H}=\left(\ket{D} - \ket{A}\right)/\sqrt{2}$ and $\ket{V} =\left(\ket{D} + \ket{A}\right)/\sqrt{2}$:
\begin{equation}
\begin{split}
\braket{\vec{r}|\bm{U}_{1}|\Psi} &= \left( \frac{a}{\sqrt{2}}\chi_{x}(x-\delta_{x})\chi_{y}(y) + \frac{b}{\sqrt{2}} \chi_{x}(x)\chi_{y}(y) \right) \ket{D} \\
&- \left( \frac{a}{\sqrt{2}}\chi_{x}(x-\delta_{x})\chi_{y}(y) - \frac{b}{\sqrt{2}}\chi_{x}(x)\chi_{y}(y) \right) \ket{A}
\end{split}
\end{equation}

Next in the measurement sequence, the second walk-off crystal implements 
$\bm{\pi}_D$. The $y$ pointer state $\ket{\chi_{y}}$ associated with the $\ket{D}$ polarization is shifted by $\delta_y$, which we describe by $\bm{U}_2$:
\begin{equation}
\begin{split}
\braket{\vec{r}|\bm{U}_{2}\bm{U}_{1}|\Psi} &= \left( \frac{a}{\sqrt{2}}\chi_{x}(x-\delta_{x})\chi_{y}(y-\delta_y) + \frac{b}{\sqrt{2}} \chi_{x}(x)\chi_{y}(y-\delta_y) \right) \ket{D} \\
&- \left( \frac{a}{\sqrt{2}}\chi_{x}(x-\delta_{x})\chi_{y}(y) - \frac{b}{\sqrt{2}}\chi_{x}(x)\chi_{y}(y) \right) \ket{A}
\end{split}
\end{equation}
Finally, the last measurement $\bm{\pi}_H$ is strong. It is implemented by a polarizing beam splitter (PBS) with transmits only $\ket{H}$ to the camera. After the PBS, the total state is given by:
\begin{equation}
\begin{split}
\braket{\vec{r}|\bm{\pi}_H\bm{U}_{2}\bm{U}_{1}|\Psi}
&= \frac{a}{2}\left( \chi_x(x-\delta_x)\chi_y(y-\delta_y)+ \chi_x(x-\delta_x)\chi_y(y) \right)\ket{H} \\
&+ \frac{b}{2}\left( \chi_x(x)\chi_y(y-\delta_y) - \chi_x(x)\chi_y(y) \right)\ket{H}.
\end{split}
\end{equation}
Thus by measuring the final state of the pointers, the camera records the probability $\mathrm{Prob}(x,y,H)$ that a photon has a position $(x,y)$ \textit{and} polarization $\ket{H}$:
\begin{equation}
\mathrm{Prob}(x,y,H)
= \left | \frac{a}{2}\left( \chi_x(x-\delta_x)\chi_y(y-\delta_y)+ \chi_x(x-\delta_x)\chi_y(y) \right) \\
+ \frac{b}{2}\left( \chi_x(x)\chi_y(y-\delta_y) - \chi_x(x)\chi_y(y) \right) \right| ^2.
\end{equation}
We can now compute the various expectation values required. For instance:
\begin{equation}
\begin{split}
\braket{\bm{x}_H\bm{y}_D}_{\mathcal{P}, H} &= \int_{-\infty}^{\infty}\int_{-\infty}^{\infty} xy \mathrm{Prob}(x,y,H)dxdy  \\
&=\frac{|a|^2}{4}\delta_x\delta_y(1 + e^{\frac{-\delta_y^2}{8\sigma_y^2}}) + \frac{1}{4}\frac{(ab^* + a^*b)}{2}(\delta_x\delta_y e^{\frac{-\delta_x^2}{8\sigma_x^2}})
\end{split}
\end{equation}
which in the weak measurement limit ($\delta_x~\ll~\sigma_x$ and $\delta_y~\ll~\sigma_y$) reduces to
\begin{equation}
\braket{\bm{x}_H\bm{y}_D}_{\mathcal{P},H} = \frac{\delta_x\delta_y}{4}\left( 2|a|^2 + \frac{ab^* + a^*b}{2} \right ).
\end{equation}
To be clear, the first subscript $H$ in $\braket{\bm{x}_H\bm{y}_D}_{\mathcal{P},H}$ indicates that the first weak measurement $\bm{\pi}_H$ is coupled to the $\bm{x}$ pointer, while the second $H$ subscript indicates that the last strong measurement is $\bm{\pi}_H$. If we repeat similar calculations for the other required expectation values (and take the Fourier transform where required), we obtain
\begin{equation}
\begin{split}
\braket{\bm{p}_{xH}\bm{p}_{yD}}_{\mathcal{P},H} &= \frac{\delta_x\delta_y}{16\sigma_x^2\sigma_y^2}\left(\frac{ab^* + a^*b}{2}\right) \\
\braket{\bm{p}_{xH}\bm{y}_D}_{\mathcal{P},H} &= \frac{\delta_x\delta_y}{8\sigma_x^2}\left(\frac{iab^* - ia^*b}{2}\right) \\
\braket{\bm{p}_{yH}\bm{x}_D}_{\mathcal{P},H} &= \frac{\delta_x\delta_y}{8\sigma_y^2}\left(\frac{-iab^* + ia^*b}{2}\right).
\label{eqn:expec_values}
\end{split}
\end{equation}
For our crystals, the pointer shifts are the same $\delta_x = \delta_y = \delta$, as well as the pointer widths $\sigma_x = \sigma_y = \sigma$. We can introduce the momentum space width $\sigma_p$ of the pointer state by noting that $\sigma \sigma_p = 1/2$ (see Eq.~\ref{eqn:pointer_pos}). With these changes, we find that:
\begin{equation}
\begin{split}
\mathrm{Re}[\rho(H,H)] &= \frac{2}{\delta^2}\left ( \braket{\bm{x}_H\bm{y}_D}_{\mathcal{P},H} - \frac{\sigma^2}{\sigma_p^2}\braket{\bm{p}_{xH}\bm{p}_{yD}}_{\mathcal{P},H} \right ) = |a|^2, \\
\mathrm{Im}[\rho(H,H)] &= \frac{2}{\delta^2} \frac{\sigma}{\sigma_p} \left ( \braket{\bm{p}_{xH}\bm{y}_D}_{\mathcal{P},H} + \braket{\bm{p}_{yH}\bm{x}_D}_{\mathcal{P},H} \right ) = 0.
\end{split}
\end{equation}

We note that, in the case of a two dimensional state, the expectation values in Eq.~\ref{eqn:expec_values} are sufficient to obtain both parameters $a$ and $b$ of the state (up to a common phase), since $\mathrm{Re}(ab^*) \propto \braket{\bm{p}_{xH}\bm{p}_{yD}}_{\mathcal{P},H}$ and $\mathrm{Im}(ab^*) \propto \braket{\bm{p}_{xH}\bm{y}_D}_{\mathcal{P},H} - \braket{\bm{p}_{yH}\bm{x}_D}_{\mathcal{P},H}$. This implies that a single measurement sequence configuration (\textit{e.g.} $\bm{\pi}_H\bm{\pi}_D\bm{\pi}_H$) is sufficient for measuring all parameters of the $d=2$ state. However, for higher dimensional states, one would need to change the first and last measurements in the sequence in order to obtain an arbitrary density matrix element. For proof-of-concept, in the experiment we change the measurement sequence to obtain the various density matrix elements even in the $d=2$ case. Other expectation values (\textit{e.g.} $\braket{\bm{x}_V\bm{y}_D}_{\mathcal{P},H}$) are not given in this example, but can be derived with a very similar calculation.

Furthermore, we also note that the density matrix element $\rho(H,H)$ (and also $\rho(V,V)$) can be found trivially by measuring the relative intensity of the state after passing through a PBS since $\braket{\psi|\bm{\pi}_H|\psi} = |a|^2$. In general, the diagonals of a density matrix can be found easily by measuring projectors in the basis of the matrix. However, we do not rely on this approach in our experiment since we aim to compare our direct measurement method with conventional methods. Furthermore, as the dimension of the space becomes large, the diagonal elements will become an increasingly small fraction of the total elements that need to be determined. Thus, the benefit of switching procedures just for these diagonal elements will decrease as the dimension increases.

\subsection{Data Acquisition Method}

\begin{figure}
\centering \includegraphics[width=0.8\textwidth]{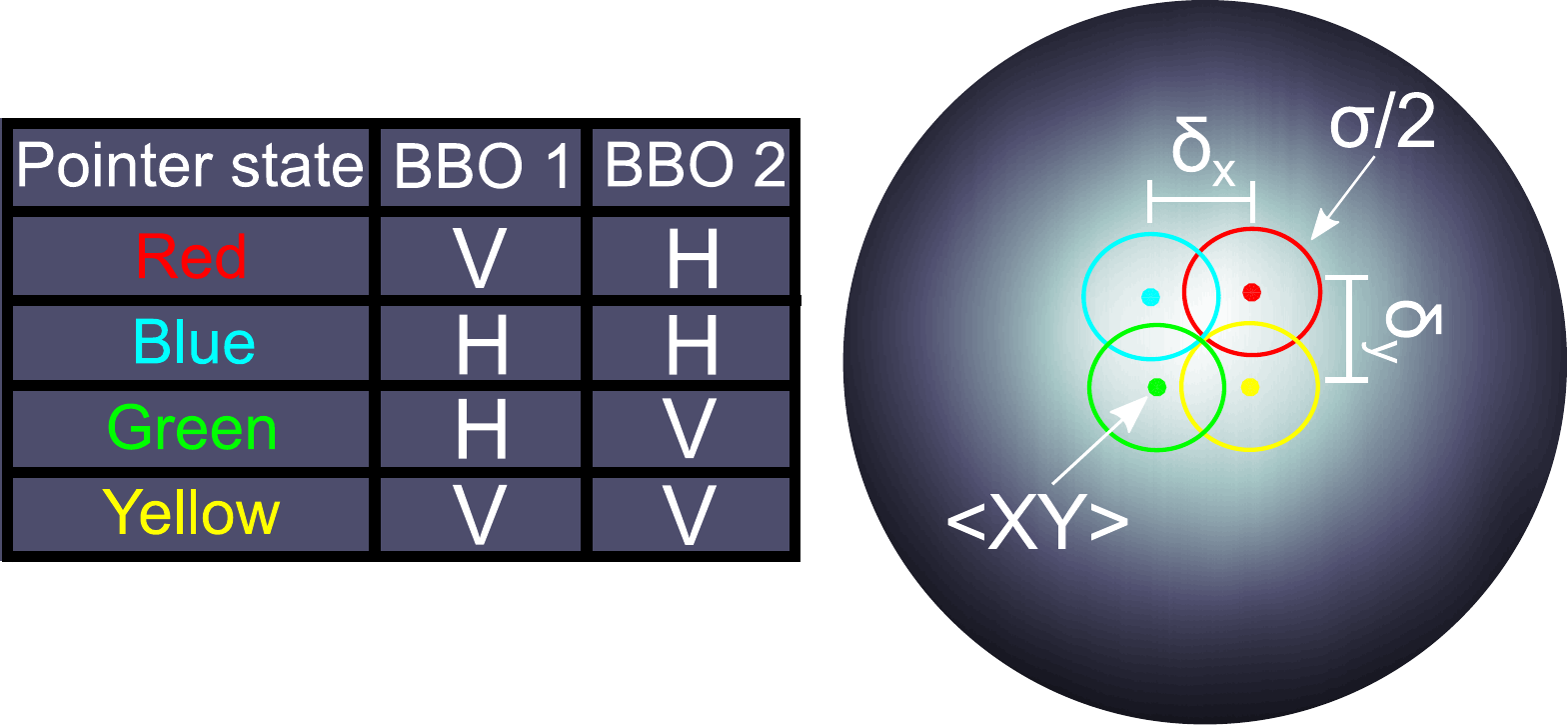} \caption{In the circle we show cropped (and superimposed) images of the four maximally shifted pointer states, after being processed. The colored circles designate the contours of the four pointer states at $\sigma/2$. The crystal shifts $\delta_x = \delta_y$ are shown for comparison. The table displays polarizations needed before each crystal to
obtain the desired pointer shift. }
\label{fig:table} 
\end{figure}

A camera captures 10 images for each input
state. Each image is processed in the following manner. First we subtract
a background image that is captured when the laser is blocked. Then
we crop the image, and apply a high-frequency Fourier filter in order
to further suppress noise contributions from the background and also
remove fringes in the image that arise due to the camera window.
Finally, we average over the 10 images to account for a random error 
of roughly $\pm$5\%
of the expectation values. This random error arises due to air fluctuations or 
small instabilities in the intensity of the laser.

The expectation values are computed with respect to the center $(x_{0},y_{0})$
of the unshifted pointer state, obtained by sending $\ket{V}$-polarized
photons through BBO 1 and $\ket{H}$-polarized photons through BBO 2,
as shown by the red circle in Fig.~\ref{fig:table}. For example,
the position expectation value is given by 
\begin{equation}
\braket{\bm{x}\bm{y}}_{\mathcal{P},J}=\sum_{i,j}(x_{i}-x_{0})(y_{j}-y_{0})\mathrm{Prob}(x_i,y_j,J)\label{eqn:expec_xy}
\end{equation}
where $\mathrm{Prob}(x,y,J)$ is a greyscale image of the pointer state
with no Fourier lens, normalized to the intensity of the
pointer state before any measurements. Similar equations as Eq.~\ref{eqn:expec_xy} are used to obtain
the other joint momentum and position expectation values.

We super-impose images of the maximally shifted pointer states in Fig.~\ref{fig:table}
to demonstrate the size of the shifts relative to the pointer widths.
The colored bold lines designate the contour of the pointer states at $\sigma/2$. The table
in Fig.~\ref{fig:table} provides the polarization states required before each crystal
to obtain the desired pointer shift. For example, the blue pointer
state is obtained when we send $\ket{H}$-polarized photons through the
first BBO crystal, then $\ket{V}$-polarized photons through the second
BBO crystal. This pointer state is maximally shifted in $x$ and we
can use it to obtain $\delta_{x}$ by computing $\delta_{x}=\braket{\bm{x}}-x_{0}$ where $\braket{\bm{x}}$ is its expectation value along $x$ and $x_{0}$
is the center of the unshifted (red) pointer state. Similarly,
the yellow pointer state is maximally shifted in $y$ and is used to
obtain $\delta_{y}$. We find that $\delta_{x}=\delta_{y}=175~\mu m$
and the pointer width is $\sigma_{x} = \sigma_{y} = 250~\mu$m (standard deviation as defined in Eq.~\ref{eqn:pointer_pos}, or 830~$\mu$m full-width at half-max). A cylindrical
and spherical lens are used to take a Fourier transform. We find that
the pointer widths in the Fourier domain are $\sigma_{p_{x}}=\sigma_{p_{y}}=90~\mu m$ (or 300~$\mu$m full-width at half-max).

\subsection{Barium Borate Crystals Alignment}

We use two birefringent barium
borate (BBO) crystals (also referred to as walk-off crystals in main text) to couple
the spatial and polarization degrees of freedom of the photons and
preform the weak measurements. To align the first BBO, we first remove
the second BBO, and send $\ket{D}$-polarized photons through the experimental
setup. The crystal axis is oriented to cause a spatial walk-off between
the respective pointer states of the $\ket{H}$ and $\ket{V}$ components
of the photons, along $x$. This achieves the projection $\bm{\pi}_H$. 
We then tilt the crystal axis along
the direction of the walk-off to compensate for the different optical
path length between each polarization component.
To verify that we aligned properly, we set the last strong projection
measurement to $\bm{\pi_A}$ in order to project onto the orthogonal polarization $\ket{A}$. 
We then adjust the tilt of the crystal until
two faint spots displaced in $x$ can be seen on the camera, 
which indicates that the pointer states of each polarization
components are interfering destructively where they overlap. A very similar
procedure is repeated to align BBO 2.

\subsection{Mixed State Generation}

\begin{figure}
\centering \includegraphics[width=0.5\textwidth]{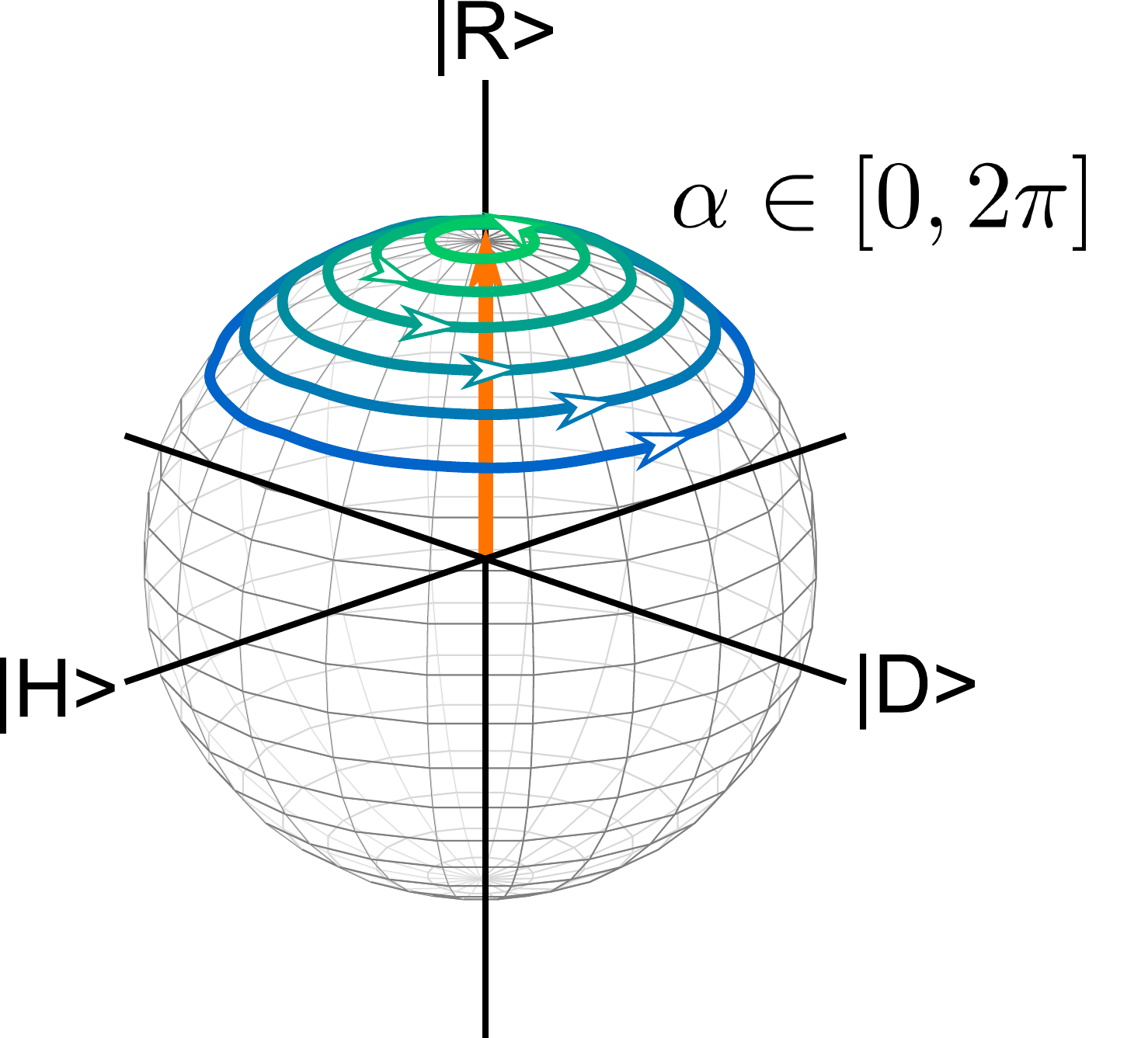}
\caption{The generation of mixed states can be visualized on the Poincar\'e sphere. Each angle $\phi$ corresponds to a certain latitude on the sphere. Over the exposure time of the camera we generate a mixture of states in $\alpha \in [0,2\pi]$ at a given latitude $\phi$, and thus the resulting state lies along the $\ket{R}$ axis, as given by Eq.~\ref{eqn:mixed_state_appendix}. In order the rotate the $\lambda / 2$ sufficiently quickly over the exposure time of the camera, we mount it on a brushless motor. }
\label{fig:poincare} 
\end{figure}

In order to generate mixed states, we create a classical mixture of pure states by using a spinning $\lambda / 2$. Given 
$\ket{H}\bra{H}$, sending it through a $\lambda/4$ with its fast-axis at an angle $\phi$ will transform it $\bm{\rho}=\ket{\psi}\bra{\psi}$ where~$\ket{\psi}=(\cos^{2}{\phi}-i\sin^{2}{\phi})\ket{H}+\frac{1}{2}(1+i)\sin{2\phi}\ket{V}$. We subsequently send this sate through a $\lambda / 2$, which is described by the unitary
$\bm{U}=\begin{pmatrix} \cos{2\alpha} & \sin{2\alpha} \\ \sin{2\alpha} & -\cos{2\alpha} \end{pmatrix}$ where $\alpha$ is the fast-axis angle of the $\lambda / 2$. If the $\lambda /2$ is spinning \textit{i.e.} $\alpha \in [0, 2\pi]$ over the exposure time of the camera, the resulting state is given by
\begin{equation}
\tilde{\bm{\rho}}=\frac{1}{2\pi}\int_{0}^{2\pi}\bm{U}\bm{\rho}\bm{U^{\dagger}}d\alpha=\begin{pmatrix}1/2 & i\sin{\phi}\cos{\phi}\\
-i\sin{\phi}\cos{\phi} & 1/2
\end{pmatrix}.
\label{eqn:mixed_state_appendix}
\end{equation}
Thus by varying $\phi$, we can produce mixed states with a varying purity: $\mathrm{Tr}[\tilde{\bm{\rho}}^2] = 1/2 + 2(\sin{\phi}\cos{\phi})^2$. This procedure can be visualized on the Poincar\'e sphere, as shown in Fig.~\ref{fig:poincare}. The camera records an image that has contributions from all the polarizations states along a latitude determined by $\phi$. Averaging over these states, the resulting state $\tilde{\bm{\rho}}$ lies along the $\ket{R}$ axis, and its purity increases with its distance from the origin. The exposure time of the camera (60 ms) sets a lower bound on the angular speed
of the $\lambda/2$ of roughly 13 rad/s. This is achieved by mounting
the wave plate on a brushless gimbal motor.